# Experimental Investigation on Particle-Laden Flows of Viscoelastic Fluids in Micro-Channels Using Optical Coherence Tomography

Kasra Amini[*,1], Gustaf Mårtensson[2], Outi Tammisola[3], Fredrik Lundell[1]

[1] FLOW and Fluid Physic Laboratory, Dept. of Engineering Mechanics, KTH Royal Institute of Technology, Stockholm, Sweden
[2] Div. of Nanobiotechnology, Dept. of Protein Science, Science for Life Laboratory, KTH Royal Institute of Technology, Solna, Sweden
[3] FLOW and SeRC (Swedish e-Science Research Centre), Dept. of Engineering Mechanics, KTH Royal Institute of Technology, Stockholm, Sweden

**ABSTRACT**
The introduction of particles to fluid flow is considered as a source of alterations in the viscosity-based behavior of the macroscopic flow field. The bilateral interactions between the solid particles and the fluid elements would both lead to changes in effective viscosity and, thereby, the velocity field. Considering the nonlinear response of non-Newtonian fluids to the local shear exerted on the bulks of fluid, the initially quasi-uniform distribution of the particles might be subject to alteration as well, due to the unbalanced force distribution on the particles. The current research investigates such particle migrations for flows of Viscoelastic Fluids (VEFs) in a straight micro-channel with a 1×3.25 mm² rectangular cross-section. Aqueous solutions of Polyacrylamide polymer in concentrations of 210 and 250 ppm have been used, where the heavy, linear, long-chain structure of the polymer introduces elasticity to the fluid. The rheological measurements are presented for the characterization of the viscoelastic behavior of the fluid samples, and the results are compared with similar flow conditions, however, in particle-laden glycerol flow as a Newtonian reference case. The flow measurements are performed using the state-of-the-art Optical Coherence Tomography (OCT) in 2D acquisition and doppler modes (D-OCT) to simultaneously resolve tomographic velocity field, and the transition of particles through the monitored cross-sections. Through the implementation of the experimental method in the current manuscript, the capability and convenience of using OCT for the problem at hand are demonstrated, as the abovementioned obtained data were to be equivalently captured by simultaneous use of Particle Image Velocimetry (PIV), for the ambient medium velocity field, and Lagrangian Particle Tracking (LPT) schemes, for identification and tracking the position of the particles. The velocity field is obtained with the spatial resolution of 2.58 µm in the depth direction, and through sub-pixel image processing, highly accurate positioning of the particles is realized. The experimental results are then used for statistical calculations, such as the Probability Distribution Function (PDF) of the cross-sectional map of the space frequented by the particles to explain the underlying physics.

**INTRODUCTION**
The interplay of viscous and elastic effects in duct flows of Viscoelastic Fluids (VEFs) results in distinct distribution of rigid particles compared to the Newtonian flow fields[1]. Parameter studies show sub-regimes for the dynamics of particle migrations based on flow/fluid geometry and properties[2]. The shear thinning effects render the field uneven in terms of local viscosity distribution as a damping force for inertial dynamics. On the other hand, elastic forces rooted from the fluid molecular structures introduce unbalanced distribution on the particle peripheries. In highly elastic flows of low inertial conditions, a particle focusing along the channel centreline occurs as the elasticity leads to the migration of particles towards the duct axis, whereas the shear thinning alongside the presence of secondary flows move the particles away from the centreline. The equilibrium, therefore, determines the eventual positioning of the particles[3].

*Corresponding author; kasraa@kth.se

From a methodological viewpoint, Optical Coherence Tomography has shown great capability in a flow field and tomographic measurements in a range of flow conditions. Near-wall errors associated with methods such as LDV and HWA[4] do not exist in OCT. There is no dependence to faithful tracer particles, and thereby particle concentration issues PIV and PTV deal with in high shear regions, such as near-wall[5], are not crucial in OCT. The Doppler OCT (D-OCT) not only work better, but also requires, a certain level of opaqueness in the media under measurement. This makes it a perfectly compatible measurement technique for non-Newtonian fluids which are mostly opaque[6]. And finally, as the illumination source is a continuous coherent light beam, there is no synchronization issues between the light source and image acquisition system, and also the ambient light noise level is second to none.

**FLUID PREPARATION AND RHEOLOGY**

As a purely Viscoelastic Fluid (VEF), aqueous solutions of Polyacrylamide PAA (FLoPAM AN934SH, SNF) have been used. The long-chain polymer is prepared by solving its dry powder in water. However, as the resulting viscoelasticity and shear thinning behaviors of the fluid is extremely dependent to the preparation process, the exact same protocol has been maintained for all samples. This protocol contains mixing at 800 rpm for 24 hours, 12 hours of rest, and 2 hours of mixing at 100 rpm for deaeration.

As is explained in the following section, for velocity field measurements, Rhodamine powder at the rheologically insignificant concentration of 0.0285% has been added and mixed with the polymer for 8 hours at 400 rpm. This serves as a contrast means required for doppler velocity measurements with OCT. All fluids are then kept at rest for a minimum of 24 hours before any measurement.

The rheometry of the fluids have been performed with Anton Paar MCR 702e. As the flow measurements are done at room temperature, the rheometer has also been set to 25°C. The concentric cylinder configuration (Bob and Cup) has been used with the external diameter of 43 mm and the internal diameter of 45 mm, leaving a 1 mm gap in between. **Figure 1** shows the result of rheometry for 210 and 250 ppm Polyacrylamide.

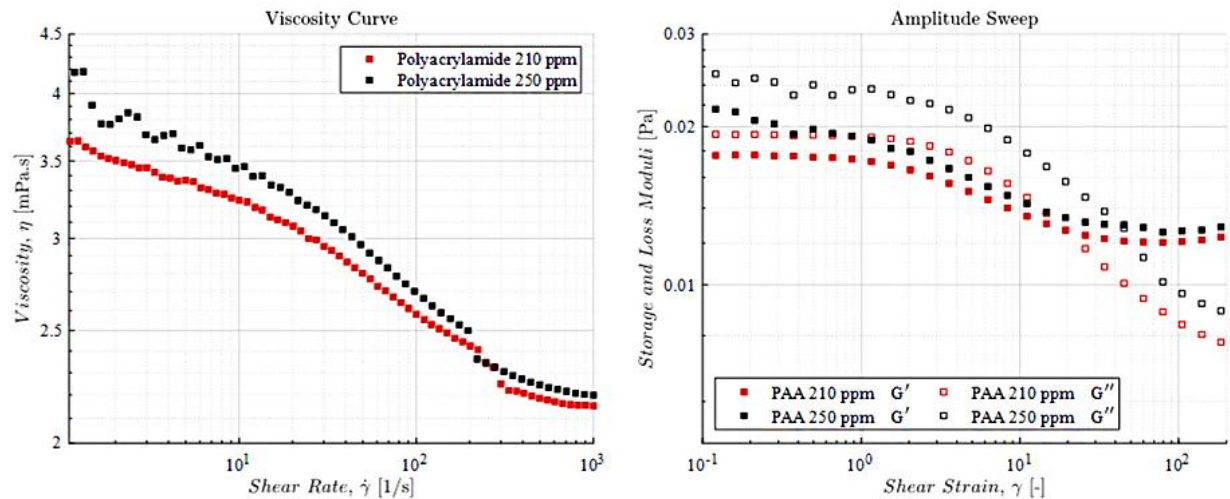

**Figure 1**. Rheological measurements of Polyacrylamide 210 and 250 ppm.

## METHODOLOGY

The test rig is a milli-fluidic straight duct with rectangular cross-section. The total length of the duct is 90 mm, with constant height of 1 mm and 3.25 mm along the axis. The fabrication of the duct, as shown in **Figure 2**, is based on a sandwich structure containing the top and bottom surfaces made from 3mm thick transparent PMMA, which are connected together with a stainless-steel plate, 1 mm in thickness, and representing the side walls of the duct pressed with the top and bottom most aluminium plates with a large number of screws. A syringe pump drives the fluid into the inlet, whereas the fluid is then drained to the atmospheric conditions at the outlet.

As the measurement device a Spectral Domain Optical Coherence Tomography (SD-OCT) Telesto II of Thorlabs has been used. The Doppler OCT (D-OCT) is the main recording mode for the current study. However, for particle-laden cases all fluids have been used in their natural state without any means of contrast. Therefore, the OCT beam does not record any velocity for the flow, other than the points/instances, where a particle crosses the plane of measurement. The intensity field captured through OCT has then been used for particle identification through image processing algorithms.

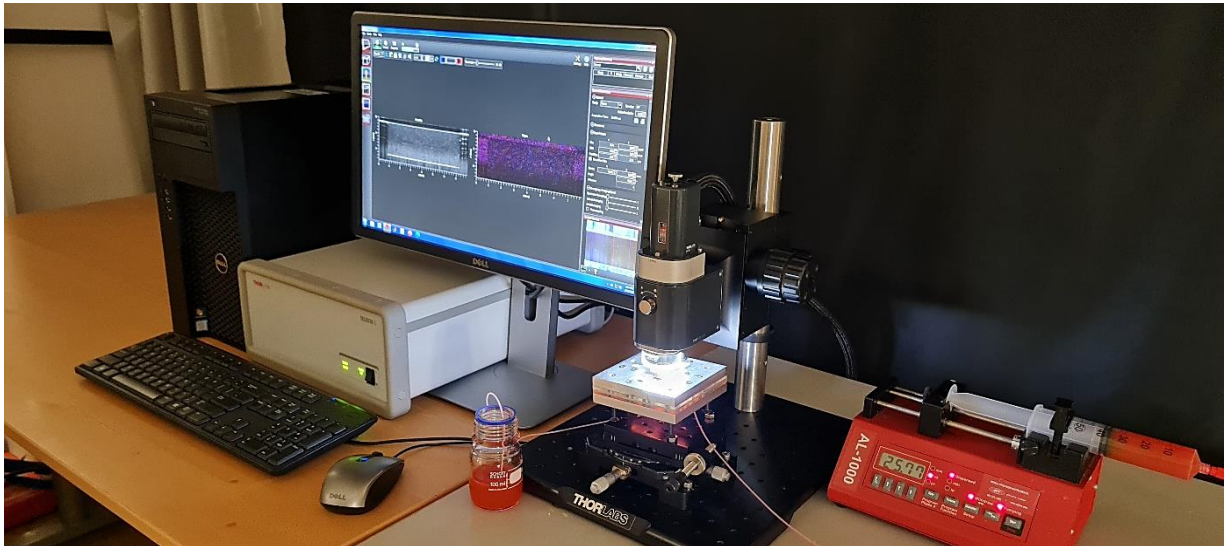

**Figure 2.** Experimental setup - OCT apparatus

Equations (1) to (5) present the formulations of the non-dimensional number groups used in this study. Reynolds number (Re) has been defined based on the half-height of the duct and the bulk velocity. The coefficient of viscosity for all calculations is based on the rheological measurements performed at the shear rates corresponding to the aforementioned parameters. The Weissenberg number (Wi) uses the relaxation time obtained by fitting the rheometric data to the Carreau model. The same parameters used for the Re are also used for Wi to calculate the flow time scales. Elastic number (El) has been used to obtain the effects of elasticity as a function of the fluid and the geometry, regardless of the imposed dynamics to the flow.

Stokes number (Stk) has been considered as the ratio between the timescale of the particle dynamics to that of the flow field, hence insights on fidelity of the particles. In a turbulent flow field, the flow timescales are defined based on the eddy size and velocity, which will adequately represent the physical time required for the flow to undergo such a trajectory. However, the conventional application of bulk velocity and half-height of a channel does not render the accurate timescale of the flow for the case at hand, since in the flow field there is no bulk of the flow passing along the mentioned distance with the bulk velocity. In this light, two distinct, however complementary, definitions of the Stk have been used; based on the axial location of the tomographic measurement along the duct axis and the bulk velocity (4), as the time required for average bulks of the flow to get to the test section after entering the inlet, and second, the duct half-height and the maximum velocity in its centreline (5), as an indicator of the wall momentum diffusion in the cross stream direction. This will lead to the Stk values ranging from $1.8\times10^{-6}$ to $9.1\times10^{-5}$ based on (4) and $4.3\times10^{-4}$ to $2.1\times10^{-2}$ based on (5). It should be mentioned that all values are well below the acceptable threshold[5] of 0.1 for validity of the fidelity premise of the particles. Consequently, the migration and focusing of the particles captured in this study are purely due to the force distribution on their peripheries, and not the incapability of the particles to follow the flow trajectories.

$$Re = \frac{\rho U_{Bulk} h}{\mu_{rheo.}} \quad (1)$$

$$Wi = \frac{\lambda U_{Bulk}}{h} \quad (2)$$

$$El = \frac{Wi}{Re} = \frac{\lambda \mu}{\rho h^2} \quad (3)$$

$$Stk = \frac{\tau_p}{\tau_f} = \frac{t_p u_0}{\ell_0} = \begin{cases} d_p^2 \frac{\rho_p}{18\mu} \Big/ \frac{\mathcal{L}_{meas.loc.}}{U_{Bulk}} & (4) \\ d_p^2 \frac{\rho_p}{18\mu} \Big/ \frac{h}{U_{max}} & (5) \end{cases}$$

Table 1 summarizes the fluid parameters and the non-dimensional numbers obtained for all cases reported in this manuscript.

Table 1. Case descriptions, fluid properties and flow conditions.

| Volumetric Flow Rate [ml/min] | | 0.4 | | | | 5 | | | | 10 | | | | 20 | | | |
|---|---|---|---|---|---|---|---|---|---|---|---|---|---|---|---|---|---|
| Bulk Velocity [mm/s] | | 2.05 | | | | 25.64 | | | | 51.28 | | | | 102.56 | | | |
| Concentration | Relaxation Time | Viscosity | Wi | Re | El | Viscosity | Wi | Re | El | Viscosity | Wi | Re | El | Viscosity | Wi | Re | El |
| [ppm] | [s] | [Pa.s] | [-] | [-] | [-] | [Pa.s] | [-] | [-] | [-] | [Pa.s] | [-] | [-] | [-] | [Pa.s] | [-] | [-] | [-] |
| Glycerol 20% | 0.00 | 2.08E-03 | 0.00 | 0.52 | 0.00 | 2.08E-03 | 0.00 | 6.47 | 0.00 | 2.08E-03 | 0.00 | 12.94 | 0.00 | 2.08E-03 | 0.00 | 25.89 | 0.00 |
| PAA 210 | 0.17 | 3.20E-03 | 0.70 | 0.32 | 2.18 | 2.80E-03 | 8.72 | 4.58 | 1.90 | 2.55E-03 | 17.44 | 10.56 | 1.65 | 2.40E-03 | 34.87 | 22.44 | 1.55 |
| PAA 250 | 0.35 | 3.60E-03 | 1.44 | 0.28 | 5.04 | 3.00E-03 | 17.95 | 4.27 | 4.20 | 2.75E-03 | 35.90 | 9.79 | 3.67 | 2.35E-03 | 71.79 | 22.91 | 3.13 |

**RESULTS AND DISCUSSIONS**

The results presented in this section are in accordance with the case juxtapositions summarized in **Table 1**. Two concentrations of Polyacrylamide, i.e. 210 and 250 ppm, have been used as the purely viscoelastic fluid. For velocity field recordings of the Newtonian fluids through 2D B-scan D-OCT, milk has been used for its adequate contrast and scattering behavior. For Newtonian cases, however in the particle-laden measurements, where the invisibility of the medium fluid to the D-OCT has been a prerequisite, Glycerol 20% has been used substitutively.

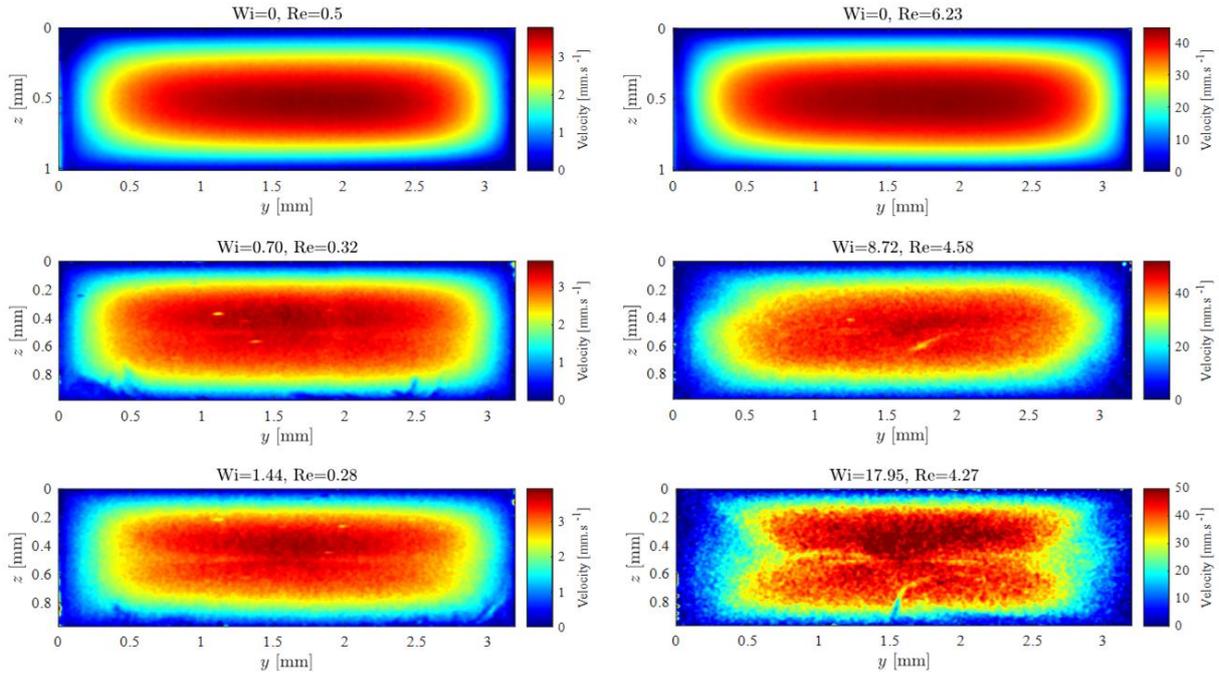

**Figure 3.** Tomographic velocity field out of cross-sectional plane for Newtonian and viscoelastic fluids.

Flow fields depicted in **Figure 3** show the comparative trend in increasing elasticity effects (Wi), in two ranges of Reynolds number. Following the left column, i.e. the low Re cases, down towards higher Wi flows, one could see the initiation of instabilities, both in central regions of the duct, where the elastic effects are strongly in play, and also propagated through the field sourcing from surface effects. Such effects of the surface are more vividly observed in the trend that right column cases take from low (Newtonian at Wi=0) to high Wi numbers. A strong asymmetry is seen in both cross-stream and streamwise planes. The maximum velocity might not occur at the center line, and the effects of the side-walls are more prominently witnessed in deviating the whole flow field away from the Newtonian paraboloid profiles prescribed by analytical solutions such as Boussinesq's solution for rectangular duct.

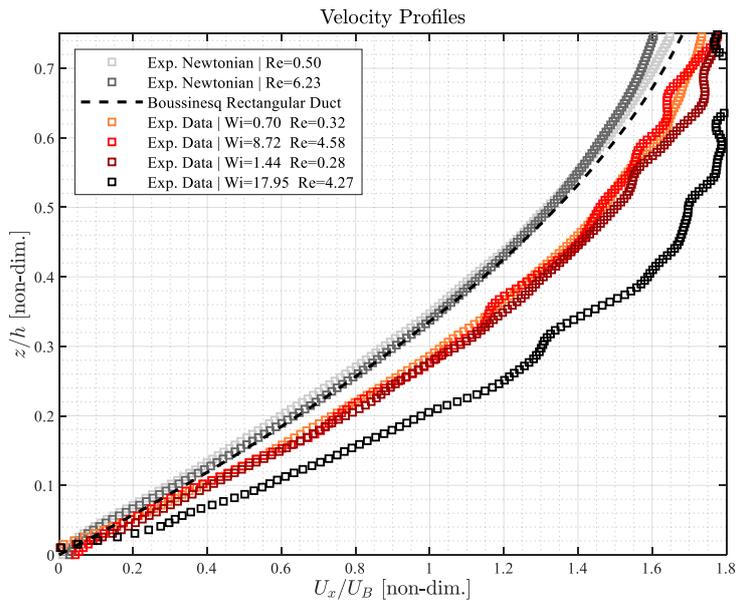

**Figure 4.** Near-wall velocity profile - Newtonian and viscoelastic fluids, and Boussinesq analytical solution for rectangular ducts.

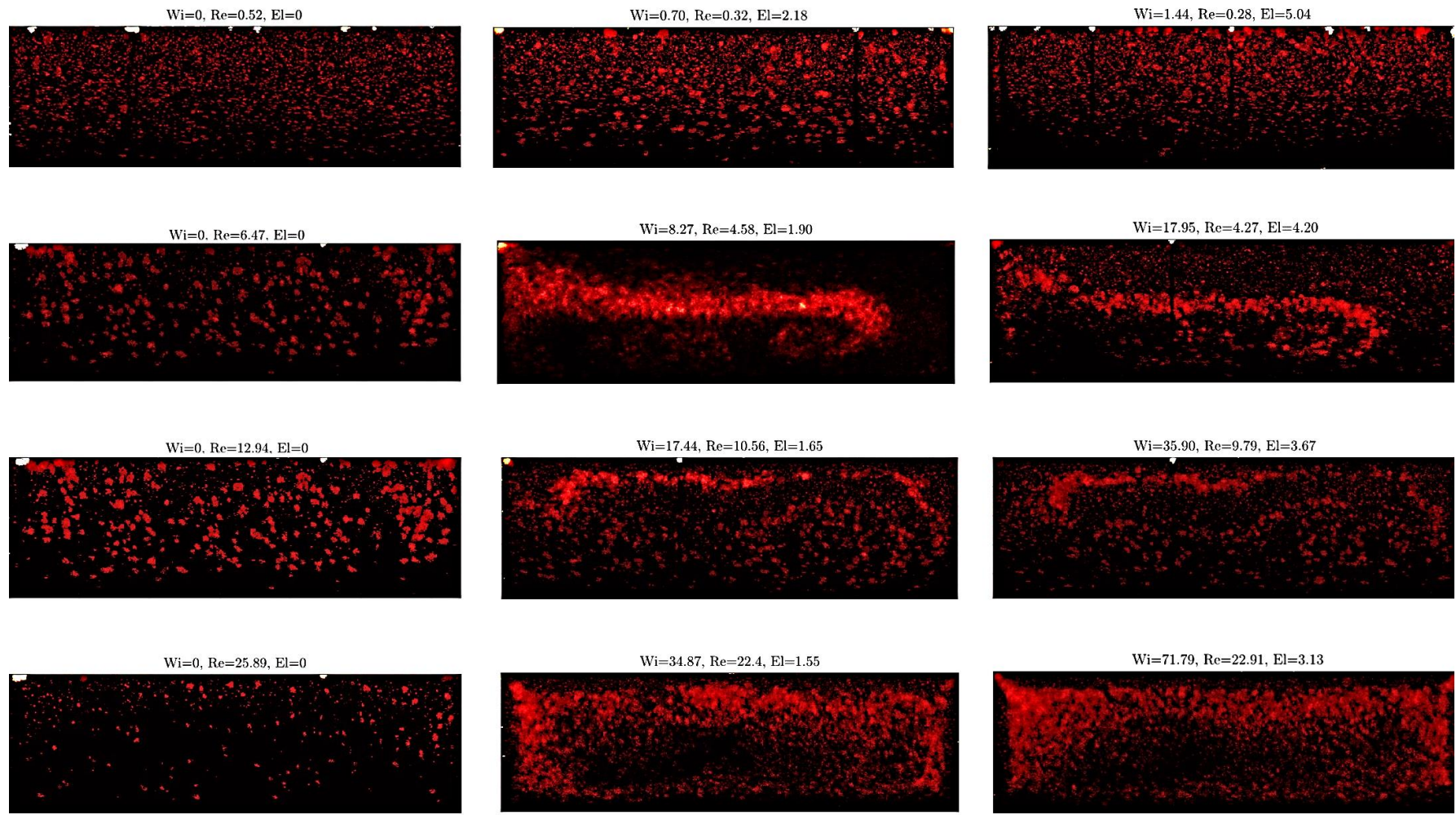

**Figure 5.** Probability Distribution Function (PDF) of the particles' migration pattern captured at 65 mm from the duct inlet. Left) Glycerol 20%, Middle) Polyacrylamide 210 ppm, and Right) Polyacrylamide 250 ppm.

**Figure 4** shows the near-wall velocity profile for all above cases. The Newtonian cases regardless of the Re show significant agreement with the analytical solution. The shear thinning effects of the Polyacrylamide leads to higher velocity gradients in near wall regions, where the highest shear rates are exerted on the flow field. The drastic fluctuations and instabilities are shown in cases with higher Wi numbers. And asymmetric velocity profiles are recognizable.

Given the described flow fields, the initially homogenous presence of rigid particles of 56 µm in diameter has been re-evaluated at the axial position of 65 mm from the duct entrance. This axial position has been chosen to fall inside the developed regions of the duct flow, and yet not too close to the outlet orifice that the flow patterns are disturbed. The particles concentration in the fluids are set to 0.8% volume fraction, to minimize the particle collisions and close interactions.

As mentioned above, the fluid media have been configured to be invisible to the OCT, so that the rigid silicon-based particles (originally used as PIV/PTV tracer particles) render the sole effects on the tomographic recordings. For each fluid/flow case 500 consecutive frames have been recorded using D-OCT to obtain a probability space for the migration of the particles. An image processing scheme identifies the particles in each frame, superimposes the location of all particles temporally in the time duration of the experiment, and obtains a tomographic Probability Distribution Function (PDF) of the particles (**Figure 5**).

Particle-laden Newtonian fluid flows have shown to maintain the quasi-uniform distribution of the particles throughout the duct. The effect of the Reynolds number is relatively weak, however the tendency of the particles to be circulated by the secondary flows in the rectangular duct is recognizable as the Re increases. However, as there is no counter-acting force in such fluids, the eventual observation is for the particles to be mixed continuously up to the axial location of the measurements. It should be noted that the lower concentration of the particles in the lower depth of the duct is solely due to the fact that the OCT beam loses its intensity as penetrating into the texture of the medium, and the particles in those regions are essentially influenced by the shadows of those flowing on top of them.

The VEFs at low Weissenberg numbers behave similarly to the Newtonian fluids. The upper row of **Figure 5** attests to this in terms of the migration of the particles, as well. As the Weissenberg number increases, however, the first stage of alterations in the distribution of the particles occurs when there is some level of elasticity, however the Re is also in the lower end of the spectrum. A focusing trend is seen as the elastic forces overcome the weak inertial forces in moving the particles towards the centreline of the duct. Keeping the Re relatively constant and increasing the Wi and thereby the El by a factor of two, a better toned and more precise focusing is noticeable in the measurements of the same axial location in the duct.

The elastic force decreases as the particles get closer to the centreline. The inertial forces, which increase with Re, tend to move the particles towards the walls. The interaction between these force groups determines the final position of the particles. In weak elastic cases, the equilibrium occurs when/where the two forces are in balance. However, a much stronger elasticity field will result in a precise focusing of the particles on the centreline.

It should be noted that the other parameters playing effective roles in the distribution of the particles are the effects of shear thinning and those of the secondary flows, both of which tend to push the particles towards the walls. Increasing the Reynolds number to ~10 shows the restructure of the particles previously oriented on the central regions of the duct towards the walls as a result of more dominant secondary flows. This pattern is more recognizable as the Re is increased up to twice the previous value, where the near wall regions are the most frequented ones of the duct cross-section. Interestingly, however, the conventional lift force exerted on the particles as they enter the regions closer to the surface compared with their dimensions, an unoccupied bound is observed on the direct vicinity of the walls.

## CONCLUSIONS

The dynamics of particle migration in milli-channel flows of VEFs is studied experimentally using Doppler Optical Coherence Tomography (D-OCT). Polyacrylamide has been used as the primary VEF, and a dilution of Glycerol has served as the Newtonian reference fluid. Spherical particles of 56 μm in diameter have been introduced to the flow homogeneously and their distribution patterns have been captured using the abovementioned tomographic technique. At low Wi and Re numbers a particle focusing occurs around the duct centreline, which is then re-oriented as the higher levels of inertia lead to the strengthened secondary flow conditions which, consequently, tend to push the particles towards the walls.

## ACKNOWLEDGEMENT

This project has received funding from European Union's Horizon 2020 research and innovation program under the Marie Skłodowska-Curie grant agreement No. 955605 YIELDGAP.